\documentclass{iau}
\usepackage{graphicx}

\title[ASAS variable stars] 
{A review of pulsating stars\\ from the ASAS data}

\author[Andrzej Pigulski]   
{Andrzej Pigulski}

\affiliation{Instytut Astronomiczny, Uniwersytet Wroc{\l}awski, Kopernika 11, 51-622 Wroc{\l}aw, Poland \\ email: {\tt pigulski@astro.uni.wroc.pl} \\[\affilskip]}

\pubyear{2014}
\volume{301}  
\pagerange{1--8}
\jname{Precision Asteroseismology}
\editors{J.A. Guzik, W.J. Chaplin, G. Handler \& A. Pigulski, eds.}
\begin{document}

\maketitle

\begin{abstract}
The All-Sky Automated Survey (ASAS) appeared to be extremely
useful in establishing the census of bright variable stars in the sky. A short
review of the characteristics of the ASAS data and discoveries based on these data and related to pulsating
stars is presented here by an enthusiastic user of the ASAS data.
\end{abstract}

\firstsection 
\section{Introduction}
The last two decades witnessed the onset of a large number of wide-field photometric surveys accomplished with
small robotic telescopes or even telephoto lenses. There were several good reasons for that: (i) the invention
of large CCD detectors, (ii) the invention of appropriate software enabling robotisation of small telescopes, (iii) the development of
wide-band Internet, and finally (iv) a good motivation for this kind of scientific research. 
As far as the latter reason is concerned, some projects were focused 
on narrow scientific topics such as a fast (on a timescale of seconds) optical follow-up of $\gamma$-ray bursts (e.g.~ROTSE; 
\cite[Akerlof et al.~2003]{Aker03}) or a detection of Kuiper Belt asteroids (TAOS; \cite[Chen 2002]{Chen02}). There were, however,
projects that from the beginning were aimed at monitoring a large part of the sky for variability. 
The idea of monitoring the whole sky for variability is very old, but
it was Prof.~Bohdan Paczy\'nski who indicated how many scientific areas can benefit from this kind of observations.
For many years he used to encourage observers to set up telescopes that would make such 
monitoring (\cite[Paczy\'nski 1997, 2000]{Pacz97,Pacz00}).
The All-Sky Automated Survey (ASAS, \cite[Pojma\'nski 1997, 2009]{Pojm97,Pojm09}) is probably the best example of a response for this call,
though other projects such as the Hungarian Automated Survey (HAT, \cite[Bakos et al.~2002]{Bako02}) can also be regarded as
Paczy\'nski's legacy.

Given the characteristics of the ASAS data (see Sect.~\ref{sect:Char}), it is obvious that they can be used mainly for the studies of
those pulsating stars that have the largest amplitudes. Out of many classes of pulsating stars, classical pulsators (all types
of Cepheids, RR Lyrae stars), high-amplitude $\delta$~Scuti (HADS) stars and Miras should be easily searched for in the ASAS data.
On the other hand --- since the project is ongoing --- the growing number of data points leads to a
lower and lower detection threshold and enables also studies of pulsating stars with lower amplitudes, like $\delta$~Scuti
stars with small amplitudes and $\beta$~Cephei stars. In this paper, we characterise the ASAS data to show their limitations
and provide a short review of the up-to-date searches for pulsating stars with these data.

\section{The ASAS project and catalogs}\label{sect:pcat}
The ASAS project started in 1996 with observations of selected fields in the southern 
sky (\cite[Pojma\'nski 1997]{Pojm1997}). The first two phases of the project,
ASAS-1 and 2,
resulted in the discovery of about 3800 variable stars (\cite[Pojma\'nski 1998, 2000]{Pojm98,Pojm00}). In 2000, the third phase 
of the project, ASAS-3, was initiated. It lasted ten seasons, from 2000 until 2009. 
With the change of CCDs in 2010, the project entered the fourth phase, ASAS-4. 
ASAS-3 and 4 cover about 70 percent of the sky, for declinations $\delta <+28^{\rm o}$ 
(\cite[Pojma\'nski 2001]{Pojm01}). A detailed description of the equipment located at Las Campanas Observatory, Chile, 
can be found in the cited papers and the 
ASAS web page\footnote{http://www.astrouw.edu.pl/asas} and will not be repeated here. 
The observations are conducted in two filters, $V$ and $I$. Starting in 2006, the project was extended to the northern hemisphere (ASAS-3 North) by placing a telescope at Maui, Hawaii. The telescope covers the northern hemisphere with a large overlap 
with observations from the southern site. Since 2006, ASAS is therefore a real all-sky survey. Recently, another project related to ASAS
has been initiated. This is the All-Sky Automated Survey for Supernovae (ASAS-SN)\footnote{http://www.astronomy.ohio-state.edu/$\sim$assassin/index.shtml} aimed at detecting bright supernovae and other transient phenomena in the entire sky.

The analysis of the acquired ASAS data resulted in the publication of three catalogs of variable stars. The first one, 
published by \cite{Pojm00}, contains nearly 3800 variable stars. It resulted from the analysis of the $I$-filter ASAS-2 data.
Variable stars in this catalog were divided into two broad categories: periodic and miscellaneous. The second catalog 
is the largest one. It is the ASAS Catalog of Variable Stars (ACVS) containing results of the analysis of the ASAS-3 $V$-filter observations. 
It was published in five parts (\cite[Pojma\'nski 2002, 2003; Pojma\'nski \& Maciejewski 2004, 2005; 
Pojma\'nski et al.~2005]{Pojm02,Pojm03,Pojm04,Pojm05,Pojma05}). The catalog contains about 50\,000 variable stars
that were classified automatically with the use of the following information: periods, amplitudes, Fourier coefficients,
2MASS colours and \textit{IRAS} infrared fluxes. Thirteen classes of variability were defined including nine for pulsating stars:
classical Cepheids, fundamental (DCEP$_{\rm FU}$), and first-overtone (DCEP$_{\rm FO}$), W~Vir stars (CW), anomalous Cepheids (ACEP), 
RR Lyrae stars of Bailey type ab (RRAB) and c (RRC), Miras (M), $\delta$~Scuti (DSCT) and $\beta$~Cephei (BCEP) stars. Multiple classifications were allowed.
It is worth noting, however, that there are also other, independent classifications of the ASAS-1/2 (\cite[Eyer \& Blake 2005]{EyBl05}) and the ACVS
(\cite[Richards et al.~2012]{Rich12}).

Finally, the third catalog is the catalog of variable stars in the \textit{Kepler} field.
It resulted from the analysis of the first 17 months of $VI$ observations by the ASAS-3 North project. 
The catalog contains only 947 stars (\cite[Pigulski et al.~2009]{Pigu09}) classified with a slightly different scheme than 
that applied in the ACVS. 

It is worth noting that both $I$-filter ASAS-2 data and $V$-filter ASAS-3 data are available not only for stars from the
ASAS catalogs of variable stars, but for all $15 \times 10^9$ observed stars, allowing different in-depth variability searches.
 
\section{Characteristics of the ASAS data}\label{sect:Char}
The way the observations are carried out in the ASAS project has a large impact on the variability we can expect to discover and
is subject to several limitations that are inherent to these data. A potential user of the ASAS data must be aware of those 
characteristics, so it is reasonable to list the most important ones. They are the following:
\begin{enumerate}
\item[\sc{Depth of the survey.}]  For the $V$-filter observations that were used to produce the ACVS, a typical exposure times lasted 3 minutes.
This allowed one to obtain reasonable photometry for stars in the $V$ magnitude range between 7 and 14, albeit some stars brighter than
$V=7$~mag and fainter than $V=14$~mag can also be found in the catalogs. The best photometry, with a typical scatter of about 0.01~mag, 
was obtained for stars in the magnitude range between 8 and 10. With almost ten years of ASAS-3 observations, the detection limit 
for periodic signals is, however, much lower than 0.01~mag. As can be judged from Fig.~16 of \cite{PiPo08b}, signals with amplitudes as
low as 4--5~mmag can be detected. This shows that variable stars with low amplitudes can also be studied with the ASAS data (see Sect.~\ref{sect:LA}).
\item[\sc{Sampling and aliasing.}] The all-sky character of the survey makes the sampling of the ASAS data quite sparse. Typically, a single field was observed 
every one, two or three days. The histogram of time intervals between consecutive observations for one of the ASAS 
targets is shown in Fig.~\ref{fig1}.
One may suppose that this type of observing would lead to problems with unambiguous determination of frequencies for stars
with periods much shorter than 1 day, e.g.~$\delta$~Scuti stars (with one-day sampling the Nyquist frequency amounts to 0.5~d$^{-1}$).
Fortunately, this is not the case for the ASAS data: not counting the alias problem (see below) there is no difficulty
in identifying a correct frequency in frequency spectra of the ASAS data. This is because the data \textit{are not exactly} evenly spaced. 
\begin{figure}[h]
\begin{center}
\includegraphics[width=\textwidth]{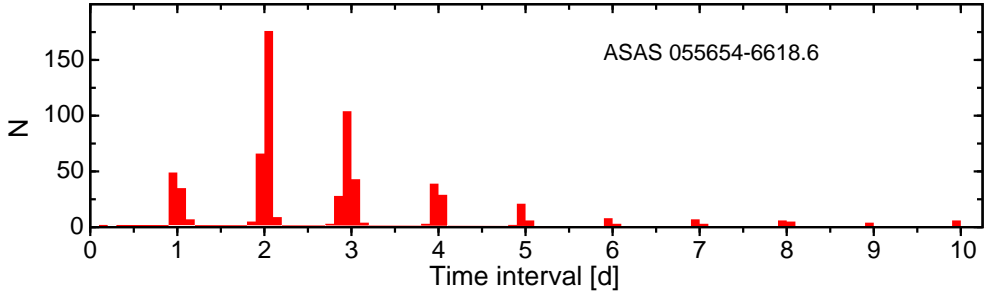} 
\caption{Histogram of time intervals between consecutive observations for one of the ASAS targets.}
\label{fig1}
\end{center}
\end{figure}
\begin{figure}[h]
\begin{center}
\includegraphics[width=0.9\textwidth]{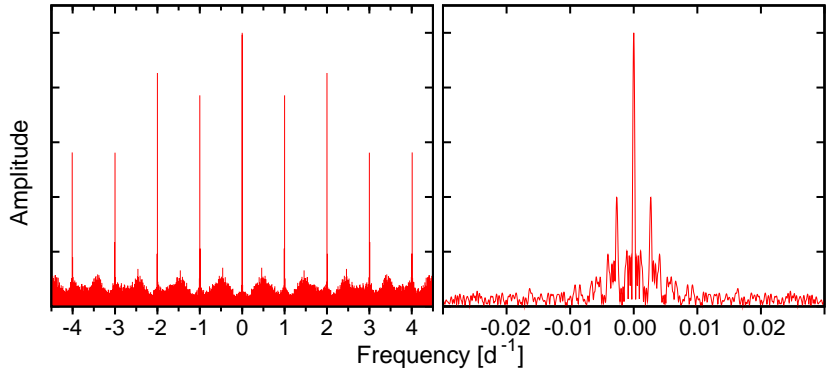} 
\caption{\textit{Left:} Window function in the frequency interval $[-4.5, 4,5]$~d$^{-1}$ for the ASAS data of the same star as in Fig.~\ref{fig1}. 
\textit{Right:} The same as in the left panel, but in the frequency interval $[-0.03, 0.03]$~d$^{-1}$ to show the yearly aliases at $\pm$0.0027~d$^{-1}$. Amplitude is given in arbitrary units.}
\label{fig2}
\end{center}
\end{figure}

The sampling shown in Fig.~\ref{fig1} results --- as one may expect --- in very strong daily aliases in Fourier spectra calculated for the ASAS data.
An example of a spectral window, for the same star as in Fig.~\ref{fig1}, is shown in Fig.~\ref{fig2}. One can easily recognize very strong daily aliases,
a consequence of the very short interval during the nights when the star was observed. On the other hand, the yearly aliases (right panel) are quite low; 
they reach about 40 percent of the height of the central peak. This is the consequence of the observing strategy: a given star was observed 
whenever possible so that usually longer gaps in data (but still lasting a few months at most) occur only for stars located close to the ecliptic.
The alias pattern seen in Fig.~\ref{fig2} has a direct consequence for
the study of short-period pulsators with the ASAS data. 
While short follow-up observations can easily resolve the problem with daily aliases we get from these data, 
the long interval they cover allows to remove ambiguities related to yearly aliases from other observations. 
An example of the advantage of using ASAS data for this purpose
is the $\beta$~Cephei-type star V403 Car (NGC 3293-16). The ASAS data helped to identify the correct frequency in this star, 
which was subject to the yearly aliasing problem (\cite[Pigulski \& Pojma\'nski 2008a]{PiPo08a}).
\item[\sc{Frequency resolution.}] As a consequence of over 9 years (3300 days) of observing within the ASAS-3 project, 
we obtain very good frequency resolution of 0.0003~d$^{-1}$ in the frequency spectra of the ASAS data. 
This is not the only advantage of the ASAS data in terms of the length of observations. Such long observations 
can be used to monitor secular changes of amplitudes
and pulsation periods. Examples can be found in the papers by \cite{PiPo08a} and \cite{Wils07}; see also Sect.~\ref{sec:Cep}.
\item[\sc{Spatial resolution.}] The large fields observed within the ASAS project resulted obviously in a poor spatial resolution.
For the equipment used to obtain ASAS-3 $V$-filter data, the scale amounted to about 15.5 arcsec per pixel. In addition, 
the ASAS magnitudes originate from aperture photometry. The data come in five different apertures with a diameter of 2--6 pixels corresponding to 0.5--1.5 arcmin on the sky. In some conditions, the problem of contamination can be resolved by analysing data in different apertures. However,
this is not always possible and only follow-up observations can solve the problem. For example, in their
search for pulsating components of eclipsing binaries using the ASAS data, \cite{PiMi07} found that the massive eclipsing O6\,V((f)) + early B-type 
binary ALS\,1135 shows, in addition to eclipses, periodic variability with a frequency of 2.31~d$^{-1}$. The frequency was tentatively attributed 
to pulsations of one of the components. Unfortunately, follow-up observations showed (\cite[Michalska et al.~2013]{Mich13}) that it is a nearby
eclipsing binary, unresolved in the ASAS frames, that is responsible for this variability.
 \end{enumerate}

\section{Pulsating stars in the ASAS data}
The algorithm used to select variable stars for the ASAS-2 and ACVS catalogs was based on the magnitude-dispersion
relation and therefore was biased toward large-am\-pli\-tude stars. However, many more variables, especially with small
amplitudes could be found using the ASAS data (see, e.g., Sect.~\ref{sect:LA} or \cite[David et al.~2013]{Davi13}).
Given the limited information used for the automatic classification of variable stars in the ASAS catalogs, no-one can expect
that the result will be unambiguous. The variability needs to be verified using other sources of information, e.g.~spectral types. 
The ASAS classification is therefore a good starting point for the study of a given class of variable stars, 
but needs to be verified.

\subsection{Cepheids}\label{sec:Cep}
The ACVS contains 1669 stars with a first classification assigned to one of the four types of Cepheids (see Sect.~\ref{sect:pcat}). 
There is no complete verification of the classification in this group, but large samples were 
included in several follow-up programs aimed mainly at the study of the structure of our Galaxy.
From multi-colour photometry (\cite[Berdnikov et al.~2009a, 2009b, 2011, Schmidt et al.~2009, Schmidt 2013]{Berd09a,Berd09b,Berd11,Schm09,Schm13}) 
and spectroscopy (\cite[Schmidt et al.~2011]{Schm11}) it became clear that genuine Cepheids amount to considerably less than a half of stars
classified as such in the ACVS. Nevertheless, the ASAS data increased considerably the number of known bright Galactic Cepheids 
and triggered several interesting projects. For example, the number of known Galactic double-mode
Cepheids was increased almost twofold with the ASAS data (e.g.~\cite[Wils \& Otero 2004, Khruslov 2009]{WiOt04,Khru09}) and includes now almost 40 members; for a complete list of Galactic double-mode Cepheids see 
the references in Sect.~3 of the paper by \cite{SmMo10}.

The ASAS data appeared also to be important in studies of the brightest (i.e.~long-period) Cepheids in the Magellanic Clouds. 
In particular, combined with archival data spanning a century or so, they were used to study secular evolutionary period changes in these Cepheids
(\cite[Pietrukowicz 2001, 2002, Karczmarek et al.~2011]{Piet01,Piet02,Karc11}). Due to the overlap in brightness with the OGLE survey, 
these observations allowed also to supplement the long-period tail of the period-luminosity relation for Cepheids 
in the Large Magellanic Cloud, only recently completed by observations from the OGLE itself (\cite[Ulaczyk et al.~2013]{Ulac13}).

\begin{figure}[!th]
\begin{center}
\includegraphics[width=0.8\textwidth]{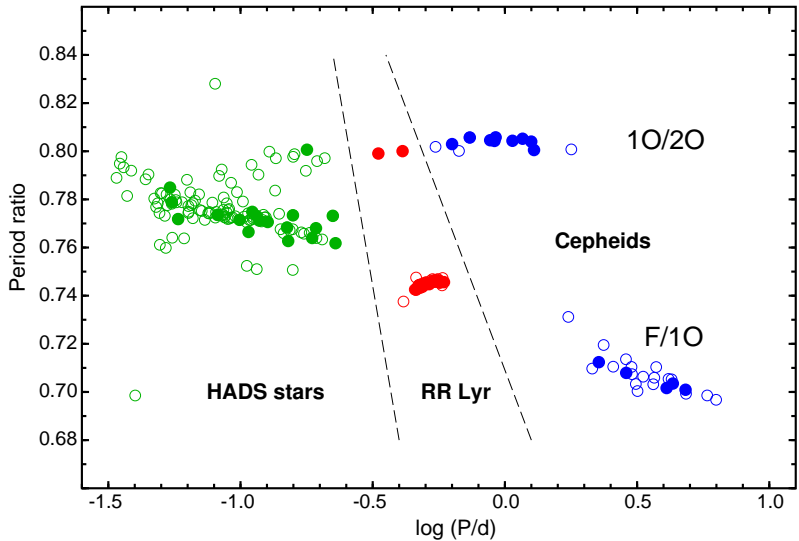} 
\caption{Petersen diagram for known double-mode Galactic classical pulsators: HADS stars (green in the on-line version), RR~Lyrae stars (red on-line), and
classical Cepheids (blue on-line). Two clear sequences of the F/1O and 1O/2O pulsators can be seen. Filled symbols denote stars that were 
discovered using the ASAS data.}
\label{fig3}
\end{center}
\end{figure}

\subsection{RR Lyrae stars}\label{sec:RR}
Of over 2200 stars classified as RR Lyrae stars in the ACVS, about 66 percent are RRab stars recognizable by their
large amplitudes and characteristic light curves with a steep ascending branch. For this reason, the automatic ACVS classification for these 
stars worked very well. This was not the case for the other group of RR Lyrae stars, of type RRc. Their light curves are more sinusoidal in shape
and for faint stars with a large scatter they are barely distinguished from very common W\,UMa-type over-contact eclipsing binaries 
(see, e.g.~\cite[Kinman \& Brown 2010]{KiBr10}). The most thorough analysis of the ASAS RR Lyrae stars was done by \cite{SzFa07} and \cite{Szcz09}. 
\cite{Szcz09} analysed only the ASAS RRab sample of stars, finding
a clear manifestation of the Oosterhoff dichotomy among them in the period-amplitude diagram.
\cite{SzFa07} focused 
on the study of multiperiodicity among these stars, finding over a hundred stars (5.5 percent of the whole sample) showing the Blazhko effect.
They also indicated a bi-modal distribution of Blazhko periods in RRc stars (see also Skarka, these proceedings). Finally,
they found four new double-mode fundamental/first overtone, (F/1O), i.e., RRd stars. At present, the total number of RRd stars found with the ASAS
data amounts to 22, i.e.~about 40 percent of all known Galactic stars of this type. Two likely first overtone/second overtone (1O/2O) RR Lyrae stars were
also found using the ASAS data (\cite[Khruslov 2010, 2012]{Khru10,Khru12}). In order to illustrate the importance of the ASAS project in studying
classical pulsators, we show in Fig.~\ref{fig3} the Petersen diagram for three classes of classical pulsators that we discuss here (Sect.~\ref{sec:Cep}\,--\,\ref{sec:HADS}).

\subsection{High-amplitude $\delta$~Scuti (HADS) stars}\label{sec:HADS}
The HADS stars form the third group of large-amplitude pulsators located in the classical instability strip. Like Cepheids
and RR Lyrae stars, they are recognised by dominating radial modes. SX Phe stars, their Population II counterparts, are strongly related to 
this group and distinguished only by much lower metallicities. Obviously, in the ACVS the 
HADS stars could be found mainly among 1275 stars classified as $\delta$~Scuti stars. 
However, we carried out our own search for these stars in the whole ACVS database. We found 148 HADS stars including 18 double-mode HADS
(Fig.~\ref{fig3}) of which eight are new discoveries. The full results of this study will be published elsewhere (Pigulski, in preparation).

\subsection{Stars with low amplitudes}\label{sect:LA}
Probably the best illustration of the fact that ASAS data can also be used to study low-amplitude pulsating stars are discoveries
related to $\beta$~Cephei stars. Of 40 stars classified as $\beta$~Cephei stars in the ACVS only 12 were confirmed as such 
(\cite[Pigulski 2005]{Pigu05}). On the other hand, six other $\beta$~Cephei stars were classified differently in the ACVS. Even
more importantly, \cite{Pigu05} recognised that all $\beta$~Cephei stars included in the ACVS have amplitudes exceeding $\sim$35~mmag
whereas a detection threshold of 5--10~mmag can be achieved. In consequence, a more thorough search for $\beta$~Cephei stars was carried
out leading to the discovery of about 300 $\beta$~Cephei stars in the ASAS-3 data (\cite[Pigulski \& Pojma\'nski 2008b, 2009]{PiPo08b,PiPo09}, 
Pigulski \& Pojma\'nski, in preparation) which means a fourfold increase in the number of known stars of this type. Five $\beta$~Cephei stars 
were also found using the ASAS-2 data (\cite[Handler 2005]{Hand05}). 

\begin{figure}[!th]
\begin{center}
\includegraphics[width=\textwidth]{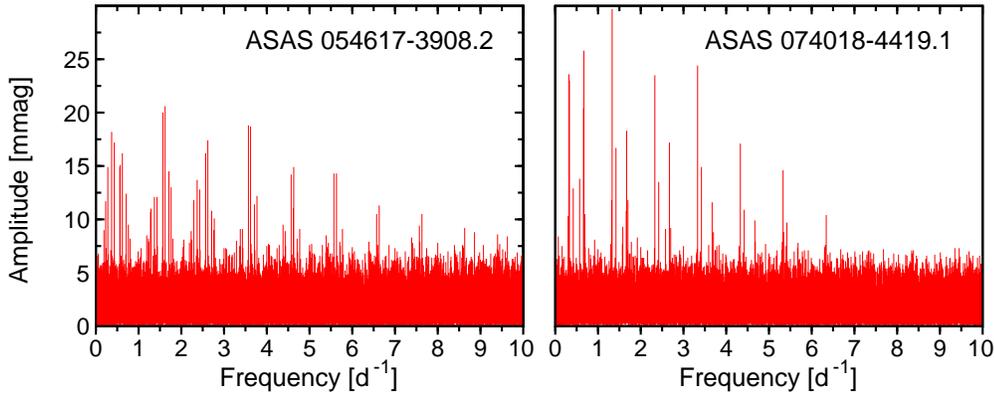} 
\caption{Frequency spectra of two $\gamma$~Doradus-type stars discovered in the ASAS data: ASAS\,054617$-$3908.3 (CD\,$-$39$^\circ$2175, left) 
and ASAS\,074018$-$4419.1 (HD\,62110, right). Four modes, with frequencies equal to 1.62557, 1.57170, 
1.71452, and 1.76774~d$^{-1}$ were detected in the former star and three (frequencies of 1.33265, 1.42350, and 1.70049~d$^{-1}$) in the latter.}
\label{fig4}
\end{center}
\end{figure}

There are other types of low-amplitude pulsating stars that were not classified within the ACVS classification scheme.
Can we detect them and study with the ASAS data?  For at least two types, slowly pulsating B (SPB) stars and $\gamma$~Doradus stars, 
the answer is obvious: yes, we can. They 
have periods ranging from about half a day to several days and amplitudes comparable to $\beta$~Cephei stars. We may therefore
expect to detect them easily. 
To illustrate this, we show frequency spectra for two $\gamma$~Dor stars discovered in the ASAS data (Fig.~\ref{fig4}).
The only complication for SPB and $\gamma$~Dor stars is strong aliasing which for frequencies below 2--3~d$^{-1}$ 
is more severe than for $\delta$~Sct or $\beta$~Cep stars due to the presence of additional aliases mirrored from negative frequencies. 

Unfortunately, there is only a small chance to detect pulsating white dwarfs, hot subdwarfs, and rapidly oscillating Ap stars in the ASAS 
data, for the following reasons.  First, the detection threshold of a few mmag for the ASAS data may be still much higher than the amplitudes observed in most of
these pulsators. This is also the case for stars showing solar-like oscillations. Second, the short periods they exhibit are comparable to the exposure time
for the ASAS-3 data (3 minutes). Due to averaging, this leads to a considerable reduction of the observed amplitudes. However, 
in favourable conditions (long periods, high amplitudes) even stars of the types mentioned above can be detected with the ASAS
data.

\section{Conclusions}
We are entering the golden age of wide-field surveys of different depth and extent that will provide an enormous amount
of photometric data suitable for different kinds of studies. The ASAS project we have briefly summarised here already showed
how the subject of pulsating stars can benefit from this type of project. With the ASAS and other all-sky surveys, 
different types of statistical studies become possible. Such surveys are also ideal for selecting best targets for follow-up studies
and targets that may help to understand different aspects of pulsations. The ASAS survey already triggered a lot of follow-up studies.
Some examples were presented even during this conference (Drobek \& Pigulski, Jurkovi\'c \& Szabados, these proceedings).
Next, the almost uninterrupted data provided by ASAS over a long time base allow for different studies of secular changes of amplitudes and periods including 
evolutionary changes or dynamical evolution and multiplicity of stars via the light-time effect (\cite[Pilecki et al.~2007]{Pile07}).

The ASAS data (also ACVS) are still unexploited and await new discoveries. We believe that digging into these data will 
bring as much fun as was experienced by the author of this note.

\acknowledgements I would like to thank Grzegorz Pojma\'nski for his cooperation in working on the ASAS data, of which I am
an enthusiastic user. This work was supported by the NCN grant 2011/03/B/ST9/02667.

\end{document}